# A Secure and Fault-tolerant framework for Mobile IPv6 based networks

Rathi S
Sr. Lecturer, Dept. of Computer Science and Engineering
Government College of Technology
Coimbatore, Tamilnadu, INDIA

Thanuskodi K
Principal,
Akshaya College of Engineering
Coimbatore, Tamilnadu, INDIA

*Abstract*— **Mobile IPv6 will be an integral part of the next generation Internet protocol. The importance of mobility in the Internet gets keep on increasing. Current specification of Mobile IPv6 does not provide proper support for reliability in the mobile network and there are other problems associated with it. In this paper, we propose "Virtual Private Network (VPN) based Home Agent Reliability Protocol (VHAHA)" as a complete system architecture and extension to Mobile IPv6 that supports reliability and offers solutions to the security problems that are found in Mobile IP registration part. The key features of this protocol over other protocols are: better survivability, transparent failure detection and recovery, reduced complexity of the system and workload, secure data transfer and improved overall performance.**

## I. INTRODUCTION

As mobile computing has become a reality, new technologies and protocols have been developed to provide mobile users the services that already exist for non-mobile users. Mobile Internet Protocol (Mobile IP) [1, 2] is one of those technologies that enables a node to change its point of attachment to the Internet in a manner that is transparent to the application on top of the protocol stack. Mobile IP based system extends an IP based mobility of nodes by providing Mobile Nodes (MNs) with continuous network connections while changing their locations. In other words, it transparently provides mobility for nodes while backward compatible with current IP routing schemes by using two types of Mobility Agents (MA), the Home Agent (HA) and the Foreign Agent (FA).

While HA is responsible for providing permanent location to each mobile user, the FA is responsible for providing Care-Of-Address (COA) to each mobile user who visits the Foreign Network. Each HA maintains a Home Location Register (HLR), which contains the MN's Home Address, current COA, secrets and other related information. Similarly, FA maintains Visitors Location Register (VLR) which maintains information about the MNs for which the FA provides services. When the MN is within the coverage area of HA, it gets the service from the HA. If the MN roams away from the coverage of HA, it has to register with any one of the FAs around to obtain the COA. This process is known as "Registration" and the association between MN and FA is known as "Mobility Binding".

In the Mobile IP scenario described above, the HAs are the single point of failure. Because all the communication to the MN is through the HA, since the Correspondent Node (CN) knows only the Home Address. Hence, when a particular HA is failed, all the MNs getting service from the faulty HA will be affected. According to the current specification of Mobile IP when a MN detects that it's HA is failed, it has to search for some other HA and recreate the bindings and other details. This lacks the transparency, since everything is done by the MN. Also, this is a time consuming process which leads to the service interruption. Another important issue is the security problem in Mobile IP registration. Since the MN is allowed to change its point of attachment, it is highly mandatory to ensure and authenticate the current point of attachment. As a form of remote redirection that involves all the mobility entities, the registration part is very crucial and must be guarded against any malicious attacks that might try to take illegitimate advantages from any participating principals.

Hence, the major requirements of Mobile IPv6 environment are providing fault-tolerant services and communication security. Apart from the above said basic requirements, the Mobile IP framework should have the following characteristics: 1) The current communication architecture must not be changed. 2) The mobile node hardware should be simple and does not require complicated calculations. 3) The system must not increase the number of times that communication data must be exchanged. 4) All communication entities are to be highly authenticated 5) Communication confidentiality and location privacy are to be ensured and 6) Communication data must be protected from active and passive attacks.

Based on the above said requirements and goals, this paper proposes "A secure and fault-tolerant framework for Mobile IPv6 based networks" as a complete system architecture and an extension to Mobile IPv6 that supports reliability and offers solutions to the registration security problems. The key features of the proposed approach over other approaches are:

better survivability, transparent failure detection and recovery, reduced complexity of the system and workload, secure data transfer and improved overall performance. Despite its practicality, the proposed framework provides a scalable solution for authentication, while sets minimal computational overhead on the Mobile Node and the Mobility agents.

## II. EARLIER RESEARCH AND STUDIES

Several solutions have been proposed for the reliability problem. The proposals that are found in [3-8] are for Mobile IPv4 and [9-15] are for Mobile IPv6 based networks. The architecture and functionality of Mobile IPv4 and Mobile IPv6 are entirely different. Hence, any solutions that are applicable for Mobile IPv4 can not be applicable for Mobile IPv6 for the reason cited here: In mobile IPv4, the single HA at the Home Link serves the MN which makes the Mobile IPv4 prone to single point of failure problems. To overcome this problem, the Mobile IPv4 solutions propose HA redundancy. But in Mobile IPv6, instead of having single HA, the entire Home Link would serve the MNs. The methods proposed in [9, 10, 11, 12] are providing solutions for Mobile IPv6 based networks.

In Inter Home Agent Redundancy Protocol (HAHA) [9], one primary HA will provide service to the MNs and Multiple HAs from different Home Links are configured as Secondary HAs. When the primary HA failed, the secondary HA will be acting as Primary HA. But, the registration delay is high and the approach is not transparent to MNs. The Home Agent Redundancy Protocol (HARP) proposed in [10] is similar to [9], but here all redundant HAs are considered from the same domain. The advantages of this approach are registration delay and computational overhead are less when compared to the other methods. But, the drawback of this approach is that the Home link is the single point of failure.

The Virtual Home Agent Redundancy Protocol (VHARP) [11, 12, 13] is similar to [10], but it deals with load balancing issues also. In [14], the reliability is provided by using two HAs in the same Home link. The primary and the secondary HAs are synchronized by using transport layer connections. This approach provides transparency and load balancing. Also, registration delay and service interruptions are less. But, if the Home Link or both HAs are failed, then the entire network will be collapsed.

Moreover, none of the above said approaches deals with the registration security even if it plays a crucial role. Registration in mobile IP must be made secure so that fraudulent registration can be detected and rejected. Otherwise, any malicious user in the internet could disrupt communications between the home agent and the mobile node by the simple expedient of supplying a registration request containing a bogus care-of-address. The secret key based authentication in Base Mobile IP is not scalable. Besides, it also can't provide non-repudiation that seems likely to be demanded by various parties, especially in commercial settings.

Many proposals are available to overcome the above said problems which can be broadly classified under the following categories: (i) Certificate Authority – Public key Infrastructure (CA-PKI) based protocol [15] (ii) Minimal public key based protocol [16] (iii) Hybrid technique of Secret and CA-PKI based protocol [17] and (iv) Self-certified public key based protocols [18].

(i) CA-PKI based mechanisms define a new Certificate Extension message format with the intention to carry information about Certificates, which now must always be appended in all the control messages. Due to high computational complexity, this approach is not suitable for wireless environment.

(ii) The Minimal Public key based method aims to provide public key based authentication and a scalable solution for authentication while setting only minimal computing on the mobile host. Even if this approach uses only the minimal public key based framework to prevent the replay attack, the framework must be executed by using complex computations due to the creation of digital signatures at the MN. This increases the computational complexity at the MN.

(iii) Hybrid technique of Secret and CA-PKI based protocol proposes the secure key combine minimal public key besides producing the communication session key in mobile node registration protocol. The drawback of this approach is found to be the registration delay. When compared to other protocols, this approach is considerably increasing the delay in registration. In addition to that, the solution to the location anonymity is only partial.

(iv) Providing strong security at the same time reducing the Registration delay and Computational complexity is an important issue in Mobile IP. Hence, for the first time self-certified public keys are used in [18, 19] which considerably reduce the time complexity of the system. But, this proposal does not address the authentication issues of CN, Binding Update (BU) messages which lead to Denial-Of-Service attack and impersonation problems.

Based on the above discussions, it is observed that a secure and fault-tolerant framework is mandatory which will tolerate inter home link failures and ensure secure registration that should not increase registration overhead and computational complexity of the system.

## III. PROPOSED APPROACH

This paper proposes a fault-tolerant framework and registration protocol for Mobile IPv6 based networks to provide reliability and security. The solution is based on interlink HA redundancy and self certified keys. The proposal is divided into two major parts: (i) Virtual Home Agent Redundancy (VHAHA) architecture design and (ii) VHAHA Registration protocol.

The first part proposes the design of fault-tolerant framework while the second part ensures the secure registration with the Mobility Agents. This proposed approach provides reliability and security by introducing extension to overall functionality and operation of current Mobile IPv6.

The advantages of this approach are: reliable Mobile IPv6 operations, better survivability, transparent failure detection and recovery, reduced complexity of the system and workload, secure data transfer and improved overall performance. The results are also verified by performing simulation. The simulation results show that with minimal registration delay and computational overhead the proposed approach achieves the desired outcome.

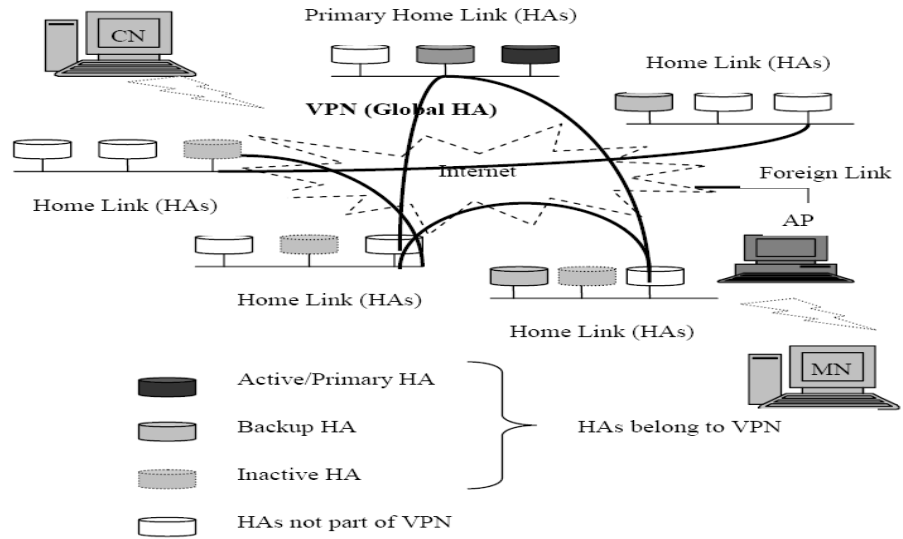

Figure 1. VHAHA Architecture

## IV. FRAMEWORK OF VHAHA

The design of VHAHA framework is divided into three major modules: (i) Architecture design (ii) VHAHA Scenario and data transmission and (iii) Failure detection and recovery algorithm.

*A.      Architecture design*

The architecture of the proposed protocol is given in Fig. 1. As part of Mobile IPv6, multiple Home Links are available in the network and each Home Link consists of multiple HAs. In this approach, one HA is configured as *Active HA*, some of the HAs are configured as *Backup HAs* and few other HAs are configured as *Inactive HAs* from the Home Link. The Active HA provides all Mobile IPv6 services, the Inactive HA provides minimal set of services and Backup HA provides mid range of services. VHAHA requires that for each MN there should be at least two HAs (one active HA and the other could be any one of the backup HA) holding its binding at any instance of time. The functionalities of these HAs are given below:

*Active HA:* There must be a HA on the Home Link serving as the Active HA. Only one HA could act as Active HA at any instance of time. The active HA maintains the *Binding cache,* which stores the mobility bindings of all MNs that are registered under it. This will hold [0-N] mobility bindings. This is responsible for data delivery and exclusive services. The *exclusive services* mainly include Home Registration, De-registration, Registration, Registration-refresh, IKE and DHAD. Besides these, it provides regular HA services such as Tunneling, reverse Tunneling, Return Routability and IPv6 neighbor discovery.

*Backup HA:* For each MN, there will be at least two HAs which will be acting as backup HAs (no limits on maximum no. of HAs). The purpose of Backup HA is to provide continuous HA services in case of HA failures or overloading. The back up HA could hold [1-N] bindings in its binding cache. This provides all the services of Active HA except the exclusive services.

*Inactive HA:* Inactive HAs will not hold any Mobility Bindings and it provides only limited services from Backup HA services since any HA in the Home Link can act as Inactive HA.

The VHAHA is configured with static IP address that is referred as Global HA Address. The Global HA address is defined by the Virtual Identifier and a set of IP addresses. The VHAHA may associate an Active HA's real IP address on an interface with the Global HA address. There is no restriction against mapping the Global HA address with a different Active HA. In case of the failure of an Active HA, the Global address can be mapped to some other backup HA that is going to act as active HA. If the Active HA becomes unavailable, the highest priority Backup HA will become Active HA after a short delay, providing a controlled transition of the responsibility with minimal service interruption. Besides minimizing service interruption by providing rapid transition from Active to Backup HA, the VHAHA design incorporates optimizations that reduce protocol complexity while guaranteeing controlled HA transition for typical operational scenarios. The significant feature of this architecture is that, the entire process is completely transparent to MN. The MN knows only the Global HA address and it is unaware of the actual Active HA. It also does not know about the transition between backup and active HAs.

*B.      VHAHA Scenario*

Two or more HAs (One active HA and minimum of one backup HA) from each Home Link are selected. Then Virtual Private Network (VPN) [20, 21, 22, 23] is constructed among the selected HAs through the existing internetworking. This VPN is assigned with Global HA address and it will act as Global HA. HAs of the VPN will announce their presence by periodically multicasting *Heart Beat* messages inside the VPN. So, each HA will know the status of all other HAs in the Private network.

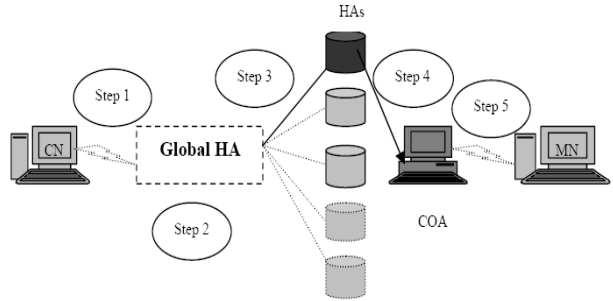

Figure 2. VHAHA Scenario.

The scenario of VHAHA protocol is given in Fig. 2. The protocol works at layer 3. In this approach, the HAs are located in different Home Links still sharing the same subnet address. The shared subnet address is known as Global HA address and the HAs in inter home link are identified by using Local HA addresses. The data destined to the MN will be addressed to the Global HA address of the MN, which will be decapsulated by the Active HA and forwarded to the MN appropriately using base Mobile IP. The various steps in forwarding the data packets are illustrated in Fig. 2.

| CN | MN | Global HA | Payload |
|---|---|---|---|

(3a) Packet sent from CN to MN

| Nearest HA | Active HA | CN | MN | Global HA | Payload |
|---|---|---|---|---|---|

(3b) Packet received by Nearest HA and directed to Active HA

| Active HA | COA | CN | MN | Global HA | Payload |
|---|---|---|---|---|---|

(3c) Packet sent from Active HA to COA of MN

| CN | MN | Global HA | Payload |
|---|---|---|---|

(3d) Packet finally received by MN

Figure 3. Packet Formats

The packet formats are shown in Fig. 3. As in Mobile IPv6, the CNs and MNs only know about the Global HA address. The packet (Fig. 3a) addressed to the MN from CN (Fig. 2. Step 1) will be directed to the Home Network using the Global HA address (Fig. 2. Step 2) of the MN. Here, the Home Network refers to the VPN that is constructed by using the above procedure. Once the packet reaches the Global HA address, all HAs that belong to Global HA address will hear the packet and the one which is closer to the MN and has less workload will pick up (Fig. 2. Step 3) the packet (Fig. 3b) using the internal routing mechanism. Then the packet will be routed to the Active HA and this Active HA will do the required processing and tunnel the packet (Fig. 3c) to the COA of the MN (Fig. 2. Step 4). Finally, the COA decapsulate and send the packet (Fig. 3d) to the MN using base Mobile IPv6 (Fig. 2. Step 5).

*C.      Failure detection and Recovery*

In contrast to Mobile IPv6 and other approaches, failure detection and tolerance is transparent to the MN. Since the MN is unaware of this process, over-the-air (OTA) messages are reduced, the complexity of the system is reduced and the performance is improved. The failure detection and recovery algorithm is illustrated in procedure 1.

___

**Begin**
**Calculate the priority for HAs that are part of Virtual Private Network**
workload($HA_i$) ← (Current mobility bindings of $HA_i$ x
    Current Throughput) / (Maximum no. of mobility
    bindings of $HA_i$ x Maximum Throughput)
Priority($HA_i$) ← 1/workload($HA_i$)

**If(HAs failed to receive heartbeats from HAx)**
    HAx ← Faulty

**If(HAx == Faulty) Then**
    Delete entries of HAx from the tables of all $HA_i$, where $1 \leq i \leq n$, $i \neq x$
  **If(HAx == Active HA) Then**
        Activate exclusive services of Backup HA
        Active HA ← Backup HA with highest priority
        Backup HA ← Select_Backup_HA (Inactive HA with highest priority),
        activate the required services and acquire the binding details from primary HA to synchronize with it
  **If(HAx == Backup HA) Then**
        Backup HA ← Select_Backup_HA(Inactive HA with highest priority),
        activate the required services and acquire the binding details from primary HA to synchronize with it.
  **If(HAx == Inactive HA) Then**
        Do nothing till it recovers, if it permanently goes off; select an Inactive HA from the Home Link of HAx
**End**
___
Procedure 1: Failure detection and Recovery

The workload of each HA in the VPN is calculated based on the number of mobility bindings associated with each HA. This workload is used for setting priority for the HAs. The priority is dynamically updated based on the changes in the number of mobility bindings. The heartbeat messages are exchanged among the HAs at a constant rate. These heartbeat messages are used for detecting the failure. When any HA fails, it will not broadcast the heartbeat message and all other HAs will not receive the heartbeats from the faulty one. Hence, the failure of the faulty HA can be detected by all other HAs that are part of the VPN.

Once the failure is detected, entry of that faulty HA will be deleted from all other HAs that are part of the Global HA

TABLE I. COMPARISON OF VHAHA WITH OTHER APPROACHES

| Metrics | MIPv6 | HAHA | HARP | VHARP | TCP | VHAHA |
|---|---|---|---|---|---|---|
| Recovery overhead | High | No | No | No | No | No |
| Fault tolerance mechanism | No | MN initiated | MN initiated | HA initiated | HA initiated | HA initiated |
| Fault tolerant Range | No | Covers entire range | Limited to Home Link | Limited to Home Link | Limited to Home Link | Covers entire range |
| Transparency | No | No | No | Yes | Yes | Yes |
| OTA messages exchanged for recovery | More | More | Less | Nil | Nil | Nil |

subnet. Then, if the faulty HA is Active HA, based on the priority of backup HAs anyone of the backup HA with the highest priority will be mapped to Active HA by activating its exclusive services. Now, the new Active HA will be the owner of the Global HA address. If the faulty HA is a backup HA then anyone of the Inactive HA will be set as the corresponding backup HA by activating the required services and acquiring binding cache entries from the Primary HA. If the Inactive HA is failed then nothing needs to be done. But if it permanently goes off, then any other HA from the link will be set as Inactive HA.

The significant feature of this approach is that the Global HA address will never change. Based on the availability of the Active HA, the Global HA address will be mapped to the appropriate backup HA. The CN and the MN would know only the Global HA address and do not know any thing about this mapping of addresses. All other internal mappings will be handled by the VHAHA's internal routing mechanism.

### D. Performance Evaluations

The proposed protocol will introduce certain amount of overhead in the network to construct the Virtual Network and to provide reliability. Hence, the performance of the proposed approach depends on two overheads: (a) Time and (b) Control message overhead. In the proposed approach, these two overheads depend on the following four factors: (1) VHAHA configuration (2) Home Registration (3) failure detection and recovery and (4) Over-the-air communication between MNs and Mobility Agents.

*1) VHAHA configuration:* The VHAHA is configured only during the initialization of the network and it will be updated only when the inactive HA fails. This happens to be a rare case, since most of the implementations will not take any action if the Inactive HA fails and let the Inactive HA to heal automatically because it will not affect the overall performance. Hence, this can be considered as one time cost and it is negligible. The Time complexity and message complexity introduced to the over all systems are negligible.

*2) Home Registration:* This factor depends on the total numbers and locations of Active, Backup and Inactive HAs that are part of VHAHA network. The registration messages include the number of messages required for the MN to get registered with the Active HA and the control messages required by the Active HA to update this information in all other backup and Inactive HAs of the MN. In the proposed approach, the Initial registration of the MN should take place with the Global HA address instead of with a single HA. Hence, this delay will be high when compared to the normal Mobile IP registration. The initial registration delay includes the time taken by the MN to get registered with the Active HA and the time taken by the Active HA to update this information in all other backup HAs.

The Time Complexity is O (D $\log^3 k$) and Message Complexity is O ($|1| + k\log^3 k$), where 'D' is the diameter of VHAHA and 'k' is number of active, backup and Inactive HAs of the MN.

*3) Failure detection and Recovery overhead:* The failure is detected when heartbeats are not received from a particular HA for a particular period of time (T). The heartbeat is actually multicasted using the multicast address. The number of heartbeats exchanged depends on 'T' and the time taken to detect the failure depends on the speed of the underlying wired network. After the failure is detected, it requires just a single message to switch over to the backup HA and the time taken is negligible.

The Time Complexity is O (D $\log^3 n$) and the Message Complexity is O ($|L| + n\log^3 n$), where 'D' is the diameter of VHAHA, 'n' is number of HAs that are part of VHAHA and 'L' represents the number of links that constitute VHAHA.

*4) Over-the-air messages:* This is very important factor because it is dealing with the air interface which is having less bandwidth. When OTA messages are increased performance of the system will be degraded. But in the proposed approach, the MN is not involved in failure detection and recovery process, so no OTA messages are exchanged during this process. The time and message complexity introduced by this factor is Nil.

From the above description, it is observed that the performance of VHAHA is directly proportional to the speed of the wired network because the proposal only involves the wired backbone operations. Actually, this is not a fair constraint because bandwidth of the network is very high thanks to the high speed and advanced networks.

### E. Simulation Results and Analysis

The proposed approach is compared with Simple Mobile IPv6, HAHA, HARP, VHARP, and TCP. The comparison results are given in Table 1. From the comparisons, it is found that VHAHA is persistent and has less overhead when compared to other approaches.

Simulation experiments are performed to verify the performance of the proposed protocol. It is done by extending the Mobile IP model given in ns-2 [24]. MIPv6 does not use any reliability mechanism; hence the time taken to detect and recover from the failure will be high. TCP, VHARP and VHAHA take almost same time to recover from the failure. This is in the case of the HAs from the same link fail. But, when the entire network fails, only the VHAHA survives. All other methods will collapse. The following parameters are used to evaluate the performance. (i) Failure detection and Recovery time when a HA fails in the Home Link (ii) Failure detection and Recovery time when entire Home Link fails (iii) Home Registration delay (iv) Packet loss (v) Number of messages exchanged during registration and (vi) Failure detection and Recovery messages. The simulation results are shown in Figures 4, 5, 6, 7, 8 and 9. The results are also compared with Mobile IPv6, TCP and VHARP to analyze the performance of the proposed approach.

*1) Failure detection and Recovery time when a HA fails in the Home Link:* When a particular HA is failed, all other HAs will not hear the heartbeat messages. When the heartbeat message from a particular HA is missed continuously for three times, then it is decided that the particular HA is a faulty HA. Once the failure is detected, the corresponding backup HA will be activated by the Recovery procedure. The failure detection and recovery time ($T_{FD-R}$) is calculated by using the equation (1).

$$T_{FD\_R} = 3T_H + prop.delayOfVHA\ HA \quad (1)$$

where $T_H$ represents the time required to hear the heartbeat messages by HAs.

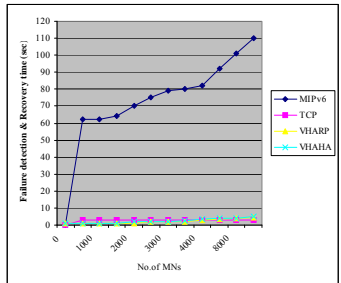

Figure 4. Comparison of Failure detection and recovery time,
when a HA fails in the Home Link

The Fig. 4 shows the $T_{FD\_R}$ of VHAHA and other protocols. Base Mobile IPv6 does not take any action for failure detection and Recovery of HAs. This needs to be handled by MN itself. Because of that, the time taken for failure detection and Recovery is very high. This causes service interruption to MNs that are affected by the faulty HA. Other schemes like TCP, VHARP and VHAHA handle the problem and almost take same amount of time for failure detection and Recovery.

*2) Failure detection and Recovery time when entire Home Link fails:* The proposed protocol constructs VPN by considering HAs from different Home links. Hence, when one Home Link fails completely also, the proposed approach handles the problem in normal manner as described in previous section. But TCP and VHARP will collapse and Failure detection and Recovery will be left to MNs.

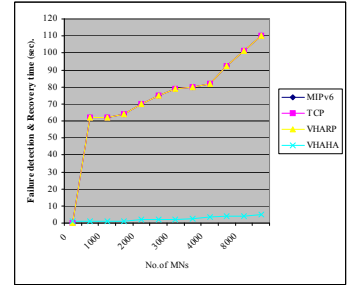

Figure 5. Comparison of Failure detection and recovery time,
when the entire Home Link fails

This situation is represented in Fig. 5, where VHAHA's Recovery time is almost equal to the previous scenario. But, TCP and VHARP approaches fail to handle the situation and Recovery time is very high which is equal to that of base MIPv6.

*3) Registration delay:* The registration delay is calculated by using the equation (2). The Active HA Registration delay is equal to that of base MIPv6. Nowadays, the bandwidth of the core network is very high and hence the propagation delay of the VHAHA is very less. The values are given in Fig. 6 and it is compared with other protocols.

$$reg.delay = reg.delay_{Active-HA} + prop.delayOfVHA\ HA \quad (2)$$

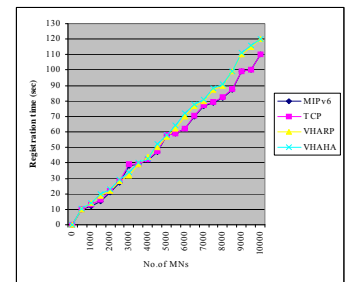

Figure 6. Comparison of Home Registration delay

*4) Packet loss:* The packet losses of the compared protocols are represented in Fig. 7. From the Figure, it is inferred that packet loss in the proposed approach is very less when compared with MIPv6, TCP and VHARP, because it is able to handle both intra link and interlink failures.

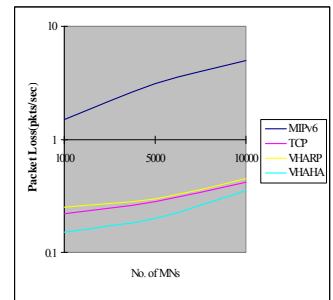

Figure 7. Comparison of Packet Loss

*5) Number of messages exchanged during registration:* This includes number of messages required to register with the Active HA and Binding Update messages to the backup HA during the Initial Registration, FA Registration and deregistration. Again, the bandwidth of the core network is very high and hence delay experienced by the MN will be negligible. This is illustrated in the Fig. 8. From the Figure, it is found that the number of messages exchanged in VHAHA is somewhat high when compared to base protocol but it is comparable with the VHARP protocol.

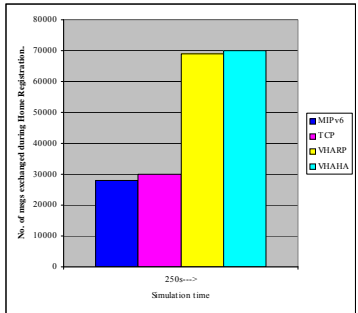

Figure 8. Comparison of no. of msgs exchanged during Home Registration

*6) Failure detection and Recovery messages:* This is represented in Fig 9. Here, also the complexity of the VHAHA is approximately equal to that of VHARP while TCP based mechanism is having less complexity and the base protocol is having the maximum complexity.

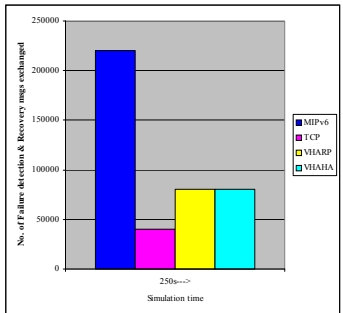

Figure 9. Comparison of no. of failure detection and Recovery messages

*F. Observation*

From the results and analysis, it is observed that the proposed approach outperforms all other reliability mechanisms because it survives even when the entire Home link fails. The overhead and complexity introduced by the proposed approach is almost negligible when compared to other existing recovery mechanisms. The failure detection and recovery overhead imposed by the proposed approach is increased by 2% when compared to VHARP. The home registration delay is also increased by 2% when compared to VHARP. The packet loss in the proposed approach is reduced by 25% when compared to all other approaches.

The template is used to format your paper and style the text. All margins, column widths, line spaces, and text fonts are prescribed; please do not alter them. You may note peculiarities. For example, the head margin in this template measures proportionately more than is customary. This measurement and others are deliberate, using specifications that anticipate your paper as one part of the entire proceedings, and not as an independent document. Please do not revise any of the current designations.

V. VHAHA SECURE REGISTRATION

The VHAHA secure registration protocol is based on self certified keys. Self-certified public keys were introduced by Girault. In contrast to the traditional public key infrastructure (PKI), self-certified public keys do not require the use of certificates to guarantee the authenticity of public keys. The authenticity of self-certified public key is verified implicitly by the proper use of the private key in any cryptographic applications in a logically single step. Thus, there is no chain of certificate authorities in self-certified public keys. This property of the self certified keys optimizes the registration delay of the proposed protocol at the same time ensuring registration security.

*A. VHAHA Secure Registration Protocol*

The proposed protocol is divided into three different parts: (i) Mobile node's initial registration with home network (ii) Registration protocol of MN (from Foreign Network) with authentication and (iii) Authentication between MN and CN.

The MN's initial registration part deals with how the MN is initially registered with its Home Network. First, the identity of the MN is verified and other details like nonce, Temporary Identity and secret key between MN and HA will be assigned to the MN.

*a. Mobile node initial registration with home network(VHAHA)*

(IR1) Verify the Identity of the MN

(IR2) Allocate nonce, Temporary ID($H(ID_{MN}//N_{HA})$ and shared secret $K_{MN-HA}$

(IR3) Transfer these details to the MN through secret channel and also store in the HA's database.

*b. Registration protocol of MN (from Foreign Network) with authentication*

*Agent Advertisement:*

(AA1) FA → MN: $M$, where $M_1$ = *Advertisement, $FA_{id}$, $MN_{CoA}$, $N_{FA}$, $w_F$*

*Registration:*

(R1) MN → FA: $M_2$, $<M_2>K_{MN-HA}$

where $M_2$ = *Request, Key-Request, $FA_{id}$, $HA_{id}$, $MN_{CoA}$, $N_{HA}$, $N_{MN}$, $N_{FA}$, $H(ID_{MN} // N_{HA})$, $w_H$*

(R2) FA: (upon receipt of R1)

- Validate $N_{FA}$ and Compute the key $K_{FA-HA}$

$$K_{FA-HA} = H_1[(w_H^{h(I_H)} + I_H)^{x_F} \mod n] = H_1[(w_H^{h(HA_d)} + HA_{id})^{x_F} \mod n]$$

- Compute *MAC*

**(R3)** FA → HA: $M_3$, $<M_3>K_{FA-HA}$, where $M_3 = M_2$, $<M_2>K_{MN-HA}$, $FA_{id}$, $w_F$

**(R4)** HA: (upon receipt of R3)
- Check whether $FA_{id}$ in $M_3$ equals $FA_{id}$ in R1.
- Compute the key,

$$K_{FA-HA} = H_1[(w_F^{h(I_H)} + I_F)^{x_H} \mod n] = H_1[(w_F^{h(FA_{id})} + FA_{id})^{x_H} \mod n]$$

- Compute MAC* and compare it with MAC value received. This is the authentication of FA by HA
- Check the identity of the MN in HA's database.
- Produce new nonce $N_{HA}'$, new Temporary ID($H(ID_{MN} \| N_{HA}')$) and new session key $K'_{MN-FA}$ and overlay the details in database.

**(R5)** HA → FA: $M_4$, $<M_4>K_{FA-HA}$

If $ID_{MN}$ is found out in HA's dynamic parameter database,
  $M_4=M_5$, $<M_5>K_{MN-HA}$, $N_{FA}$, $\{K_{MN-FA}\}K_{FA-HA}$

Else,
  $M_4 = M_5$, $<M_5>K^0_{MN-HA}$, $N_{FA}$, $\{K_{MN-FA}\}$, $K_{FA-HA}$
  $M_5 = Reply$, $Result$, $Key\text{-}Reply$, $MN_{HM}$, $HA_{id}$, $N'_{HA}$, $N_{MN}$

**(R6)** FA: (upon receipt of R5)
- Validate $N_{FA}$
- Validate $<M_4>K_{FA-HA}$ with $K_{FA-HA}$. This is the authentication of FA to HA.
- Decrypt $\{K_{MN-FA}\}K_{FA-HA}$ with $K_{FA-HA}$ and get the session key $K_{MN-FA}$

**(R7)** FA → MN: $M_5$, $<M_5>K_{MN-HA}$

**(R8)** MN: (upon receipt of R7)
- Validate $N_{MN}$
- Validate $<M_5>K_{MN-HA}$ with the secret key, $K_{MN-HA}$ used in R1. This is the authentication of MN to HA.

### c. Authentication between MN and CN

**(A1)** MN → HA₂: $M_1$ $<M_1>K_{MN-HA2}$, where $M_1$=Auth-Request, $MN_{COA}$, $CN_{COA}$, $N_{MN}$, $w_{MN}$

**(A2)** HA₂: (Upon receipt of A1)
- Validate $N_{MN}$, and compute *MAC*
- HA₂ → CN: $M_2$ $<M_2>K_{HA2-CN}$, where $M_2$= $M_1$ $<M_1>K_{MN-HA2}$

**(A3)** CN → MN: $M_3$ $<M_3>K_{CN-MN}$, $w_{CN}$, where $M_3$ =Auth-Response, $MN_{COA}$, $CN_{COA}$, $h(N_{MN})$
- Validate *MAC* and *nonce*. This is the authentication of $HA_2$ and $MN$ by $CN$.
- Compute $K_{CN-MN}$

**(A4)** MN: (Upon receipt of A3)
- Validate *MAC* and *nonce*. This is the authentication of $HA_2$ and $CN$ by $MN$.
- Compute $K_{CN-MN}$

---

Procedure 2: VHAHA Secure Registration Protocol

The second part deals with how the MN is registered with the Foreign Network when it roams away from the Home Network. There is no change in Agent advertisement part except that the MN authenticates the FA using its witness value. And, in Registration part, instead of passing the MN's actual identity, it is combined with nonce and then hashed. This provides the location anonymity. Also, the witness value is passed which enables the calculation of shared secrets.

The third part deals with the authentication between MN and CN. This authentication enables the MN to communicate with the CN directly which resolves the triangle routing problem. First MN sends the authentication request to the Home Agent (HA₂) of the MN. There the HA₂ verifies and authenticates the MN and forward the message to CN. The CN validates the MN, calculates the shared secret and sends response to MN. Finally, the MN calculates the shared secret and validates the CN. Then, the MN and CN can directly communicate each other. The details of the proposed protocol are summarized in procedure 2.

### B. Performance Evaluations

In this section, the security aspects of the proposed protocol are analyzed. The following attributes have been considered for the analyses are: (i) Data confidentiality, (ii) Authentication, (iii) Location anonymity and synchronization and (iv) Attack prevention.

*1) Confidentiality:* Data delivered through the Internet can be easily intercepted and falsified. Therefore, ensuring confidentiality of communication data is very important in Mobile IP environment. The data confidentiality of the various protocols and the proposed one is listed in the Table 2.

TABLE II.  DATA CONFIDENTILAITY ANALYSIS

| Methods | MN-FA | FA-HA | MN-HA | CN-MN |
|---|---|---|---|---|
| **Secret key** | No | No | Yes | Yes |
| **CA-PKI** | No | No | Yes | Yes |
| **Minimal Public key** | Yes | No | Yes | Yes |
| **Hybrid** | Yes | Yes | Yes | Yes |
| **Self certified** | Yes | Yes | Yes | Yes |
| **VHAHA secure Registration** | Yes | Yes | Yes | Yes |

The proposed approach achieves data confidentiality between all pairs of network components. From the table, it is found that Hybrid and Self certified approaches also provide the same result. But the computational complexities of these protocols are high when compared to the proposed one, due to the usage public keys and dynamically changing secret keys.

*2) Authentication:* Prior to data delivery, both parties must be able to authenticate one another's identity. It is necessary to avoid any bogus parties from sending unwanted messages to the entities. The Mobile IP user authentication protocol is different from the general user service authentication protocol. Table 3 shows the authentication analysis of various protocols with the proposed one. From the analysis, it is found that the VHAHA secure registration excels all approaches because it provides authentication between all pairs of the networking nodes.

TABLE III.  AUTHENTICATION ANALYSIS

| Methods | MN-FA | FA-HA | MN-HA | MN-CN |
|---|---|---|---|---|
| Secret key | None | None | MAC | None |
| CA-PKI | Digital Signature | Digital Signature | Digital Signature | None |
| Minimal Public key | None | None | Digital Signature | None |
| Hybrid | None | Digital Signature | Symmetric Encryption | None |
| Self certified | None | MAC (Static/ dynamic key) | MAC (dynamic key) | None |
| VHAHA secure Registration | TTP | MAC (Static/ dynamic) | MAC (dynamic key) | MAC (dynamic key) |

*3) Location Anonymous and Synchronization:* The proposed approach uses temporary identity instead of the actual identity of the MNs. Since, the actual location of the MN is not revealed to the outsides environment (i.e. CNs and Foreign links). Similarly, the proposed approach maintains two databases: (i) Initial parameter base and (ii) Dynamic parameter base. These are used for maintaining synchronization between MNs and HAs. The results are given in table 4.

TABLE IV.  LOCATION ANONYMITY AND SYNCHRONIZATION

| Methods | Location anonymity | Synchronization |
|---|---|---|
| Secret key | No | No |
| CA-PKI | No | No |
| Minimal Public key | No | No |
| Hybrid | No | No |
| Self certified | No | No |
| VHAHA secure Registration | Yes | Yes |

*4) Attack Prevention:* The following attacks are considered for the analysis: (i) Replay attack (ii) TCP Splicing attack (iii) Guess attack (iv) Denial-of-Service attack (v) Man-in-the-middle attack and (vi) Active attacks. Table 5 shows the attack prevention analyses of the various approaches.

TABLE V.  ATTACK PREVENTION ANALYSIS

| Methods | Re-play | Splic-ing | Guess | DOS | Man in middle | Acti-ve |
|---|---|---|---|---|---|---|
| Secret key | No | No | No | No | Yes | Yes |
| CA-PKI | No | Yes | No | Yes | Yes | Yes |
| Minimal Public key | Yes | No | No | Yes | Yes | Yes |
| Hybrid | Yes | Yes | Yes | Yes | Yes | Yes |
| Self certified | Yes | Yes | Yes | Yes | Yes | Yes |
| VHAHA secure Registration | Yes | Yes | Yes | Yes | Yes | Yes |

*C. Simulation Results and Analysis*

The system parameters are shown in Table 6. The cryptography operation time on the FA, HA and MN is obtained from [25]. The following parameters are used for the evaluation: (i) Registration delay and (ii) Registration Signaling traffic.

*1) Registration delay:* The registration delay plays an important role in deciding the performance of the Mobile IP protocol. To strengthen the security of the Mobile IP registration part, the data transmission speed can not be compromised because it will cause the direct impact on the end user. If the delay is high, then the interruption and packet loss will be more. Due to the properties of public keys, naturally the registration delay of these protocols is very high and the packet loss is also high. But certificate based protocols are not based on public keys and thanks to the properties of the certificates, the delay is less. The registration time is calculated by using the equation (3) and the results are given in table 7.

$$\mathrm{Re}\,g.Time = RREQ_{MN-FA} + RREQ_{FA-HA} + RREP_{HA-FA} + RREP_{FA-MN} \quad (3)$$

where, $RREQ_{MN-FA}$ is the time taken to send the registration request from MN to FA, $RREQ_{FA-HA}$ is the time taken to forward the registration request from FA to HA, $RREP_{HA-FA}$ is the time taken to send the registration reply from HA to FA and $RREP_{FA-MN}$ is the time taken to forward the registration reply from FA to MN.

TABLE VI.  COMPARISON OF REGISTRATION DELAY

| Methods | RREQ MN-FA (1) | RREQ FA-HA (2) | RREP HA-FA (3) | RREP FA-MN (4) | Delay (ms) (1)+(2)+(3)+(4) |
|---|---|---|---|---|---|
| Secret key | 2.7191 | 1.004 | 1.0144 | 2.7031 | 7.4406 |
| CA-PKI | 7.6417 | 5.9266 | 6.3170 | 7.6457 | 27.5312 |
| Minimal Public key | 2.8119 | 0.9966 | 10.8770 | 7.7466 | 22.4322 |
| Hybrid | 2.7934 | 16.0565 | 15.011 | 2.8007 | 36.6625 |
| Self certified | 3.4804 | 14.2649 | 1.0176 | 2.8402 | 21.6023 |
| VHAHA Registrat-ion | 3.3813 | 7.64708 | 1.0156 | 2.7615 | 14.8056 |

*2) Registration Signaling Traffic:* The computation overhead depends on the amount of traffic (i.e. the packet size) that is to be transmitted to successfully complete the registration. If the amount of signaling traffic is high means, computational complexity at the MN and the mobility agents will be high. The signaling traffic of the various protocols considered for comparison are computed and given in Table 8. From the table, it is observed that VHAHA secure registration is having the lowest traffic. Hence complexity is less both at MNs and Mobility Agents. Because of the lowest traffic, the bandwidth consumption is comparatively less.

TABLE VII. COMPARISON OF REGISTRATION TRAFFIC

| Methods | MN-FA | FA-HA | HA-FA | FA-MN | Size (bytes) |
|---|---|---|---|---|---|
| Secret key | 50 | 50 | 46 | 46 | 192 |
| CA-PKI | 224 | 288 | 64 | 128 | 704 |
| Minimal Public key | 178 | 178 | 174 | 174 | 704 |
| Hybrid | 66 | 578 | 582 | 66 | 1292 |
| Self certified | 226 | 404 | 124 | 70 | 824 |
| VHAHA Registration | 206 | 364 | 108 | 54 | 732 |

*D.    Observation*

The proposed approach does not affect the complexity of the initial registration. But, foreign network registration delay is significantly increased due to the application of security algorithms. From the procedure 2, it is understood that the proposed scheme does not change the number of messages exchanged for the registration process. But, the size of the message will be increased due to the security attributes that are passed along with the registration messages.

From the results and analysis, it is observed that the VHAHA secure registration reduces the registration delay overhead by 40% and signaling traffic overhead by 20% when compared to other approaches.

## VI. CONCLUSION AND FUTURE WORK

This paper proposes a fault-tolerant and secure framework for mobile IPv6 based networks that is based on inter home link HA redundancy scheme and self-certified keys. The performance analysis and the comparison results show that the proposed approach has less overhead and the advantages like, better survivability, transparent failure detection and recovery, reduced complexity of the system and workload, secure data transfer and improved overall performance. Moreover, the proposed approach is compatible with the existing Mobile IP standard and does not require any architectural changes. This is also useful in future applications like VoIP and 4G. The formal load balancing of workload among the HAs of the VPN is left as future work.